\documentstyle[proceedings,epsf]{crckapb}

\def\eq#1 {\begin{equation} #1 \end{equation}}
\def\mic    {\hbox{$\mu$m}}
\def\about  {\hbox{$\sim$}}
\def\ga     {\mathrel{\hbox{\raise.3ex\hbox{$>$}\llap
                                {\lower.8ex\hbox{$\sim$}}}}}
\def\la     {\mathrel{\hbox{\raise.3ex\hbox{$<$}\llap
                                {\lower.8ex\hbox{$\sim$}}}}}
\def\Ref {\bibitem[]{}}

\begin{opening}
\title{IR EMISSION FROM DUSTY WINDS --- SCALING AND \protect\\
       SELF-SIMILARITY PROPERTIES}

\author{Moshe Elitzur}
\author{\v{Z}eljko Ivezi\'{c}}
\institute{Department of Physics and Astronomy\\
           University of Kentucky\\
           Lexington, KY 40506-0055, USA}
\end{opening}

\runningtitle{Scaling of IR Emission from Dusty Winds}

\begin{document}

\begin{abstract}   Infrared emission from radiatively heated dust possesses
general scaling properties. The spectral shape is independent of overall
luminosity when the inner boundary of the dusty region is controlled by dust
sublimation; the only relevant property of the heating radiation is its
frequency profile.  For a given type of dust grains, the emission from red
giants and supergiants is essentially controlled by a single parameter --- the
overall dust optical depth.  This leads to tight correlations among different
spectral properties that explain many available observations and enable
systematic studies of large data bases.

\end{abstract}

\vspace*{-0.5cm}

\section{Introduction}

The radiation received on earth from late-type stars has undergone significant
processing in the surrounding dust shell.  Interpretation of the observations
therefore necessitates considerable theoretical effort, involving detailed
radiative transfer calculations.  These calculations have traditionally
required a large number of input parameters that fall into three categories:

\paragraph{The Star.}  Stellar input properties include the mass $M_\ast$,
luminosity $L_\ast$, temperature $T_\ast$ and the mass-loss rate $\dot M$.

\paragraph{The Shell.}  This is described by a density profile $\rho(r)$,
defined between some inner and outer radii $r_1$ and $r_2$, respectively.

\paragraph{Dust Properties.}  The dust abundance can be expressed via the
dust-to-gas ratio $\rho_d/\rho$.  The properties of individual grains must be
specified, too, and quantities widely employed include the grain size $a$, the
solid density $\rho_s$, sublimation temperature $T_{sub}$ and the absorption
and scattering efficiencies $Q_{abs}$ and $Q_{sca}$.

\paragraph{} Once these input properties are prescribed, they are plugged into
a detailed radiative transfer calculation whose output can be compared with
the observations.  The most widely used output quantity is the spectral shape
\eq{
                        f_\lambda = {F_\lambda \over F},
}
where $F_\lambda$ and $F$ are the observed flux density and bolometric flux,
respectively. Another quantity that can be compared with observations when the
angular resolution is sufficiently high is the surface brightness.

The rather large number of input parameters that had to be specified in
traditional calculations creates two major practical problems.  First, the
volume of parameter space that must be searched to fit a given set of
observations can become prohibitively large. Second, and more serious, even
when a successful fit is accomplished, its uniqueness is questionable and the
model parameters cannot be trusted as a reliable indication of the source
actual properties.

\section{Scaling}

Much of the input required in past calculations is in fact redundant.  In
Ivezi\'{c} \& Elitzur 1995 (IE95 hereafter) we were able to show that the IR
emission from late-type obeys general scaling properties that greatly reduce
the number of input parameters required to fully specify the radiative
transfer problem.  The primary input involves only the dust, for which we must
specify two types of properties: (1) Optical properties, specified through the
normalized spectral shape of the extinction efficiency $q_\lambda =
Q_\lambda/Q_{\lambda_0}$, where $\lambda_0$ is some arbitrary fiducial
wavelength; this spectral shape is controlled by the grain chemistry and
size.  (2) The overall optical depth $\tau^T_{\lambda_0}$; this is controlled
primarily by the mass-loss rate.  Additional input properties have only
secondary significance and involve the dust sublimation temperature,
$T_{sub}$, the relative thickness of the dust shell $r_2/r_1$ and the stellar
temperature $T_\ast$.  Note that $T_\ast$ is the only stellar property that
enters (and only in a secondary role).  In particular, the stellar luminosity
is irrelevant.

The proof of scaling is quite simple.  The shell inner radius $r_1$ is
controlled by dust sublimation, namely, $T_d(r_1) = T_{sub}$. Introduce the
dimensionless radial distance $y = r/r_1$. Then the radiative transfer equation
becomes
\eq{
      {dI_\lambda \over dy} = \tau^T_{\lambda}\eta(y)(S_\lambda - I_\lambda)
}
for $y \ge 1$. In this form, the equation contains no reference to either
densities or dimensions.  The only scale is determined by $\tau^T$, geometrical
quantities enter only through the dimensionless radius $y$.  The density enters
only through $\eta = n(y)/\!\int n(y)dy$, its dimensionless, normalized radial
profile.  Because the outflows around late-type stars are controlled by
radiation pressure on dust grains, the profile $\eta$ is uniquely determined by
$\tau^T$ when the effects of gravity and dust drift are negligible (IE95).  In
fact, under these circumstances, the analytic expression
\eq{
      \eta \propto {1\over y^2}\, \sqrt{{y \over y - 1 + (v_1/v_\infty)^2}}
}
provides an excellent approximation to the actual density profiles we find in
our detailed numerical calculations.  This function requires as additional
input the ratio of initial to final velocity, $v_1/v_\infty$.  However, this
parameter has a negligible effect on the result for its typical values, $\la$
0.1.

Additional input required is the stellar radiation.  This can be specified by
its flux density $F_{\ast\lambda} = F_\ast\times f_{\ast\lambda}$, where
$F_\ast$ is the bolometric flux. Each value of $F_1 = F_\ast(y{=}1)$ uniquely
determines a corresponding value of $T_1$, the dust temperature at $y = 1$, so
the reverse is also true.  And since the inner boundary is controlled by dust
sublimation, $T_{sub}$ fixes the value of $T_1$ and through it of $F_1$.  When
the stellar luminosity varies, the shell adjusts its inner boundary so that
$F_1$ and $T_1$ (= $T_{sub}$) stay the same.  The radiative transfer equation
is oblivious to these changes because the inner boundary always corresponds to
$y$ = 1.

These scaling properties can be generalized to arbitrary geometries and
density distributions (Ivezi\'{c} \& Elitzur 1996). The spectral shape of dust
IR emission is independent of overall luminosity when the inner boundary of
the dusty region is controlled by dust sublimation; the only relevant property
of the heating radiation is its spectral shape.  Densities and geometrical
dimensions are likewise irrelevant; they enter only through one independent
parameter, the overall optical depth. The geometry enters only through angles
and aspect ratios.  Dust properties enter only through dimensionless,
normalized distributions that describe the spatial variation of density, and
the wavelength dependence of scattering and absorption efficiencies.

\begin{figure}[ht]
\centering \leavevmode
\epsfxsize=\hsize \epsfbox[135 415 476 652]{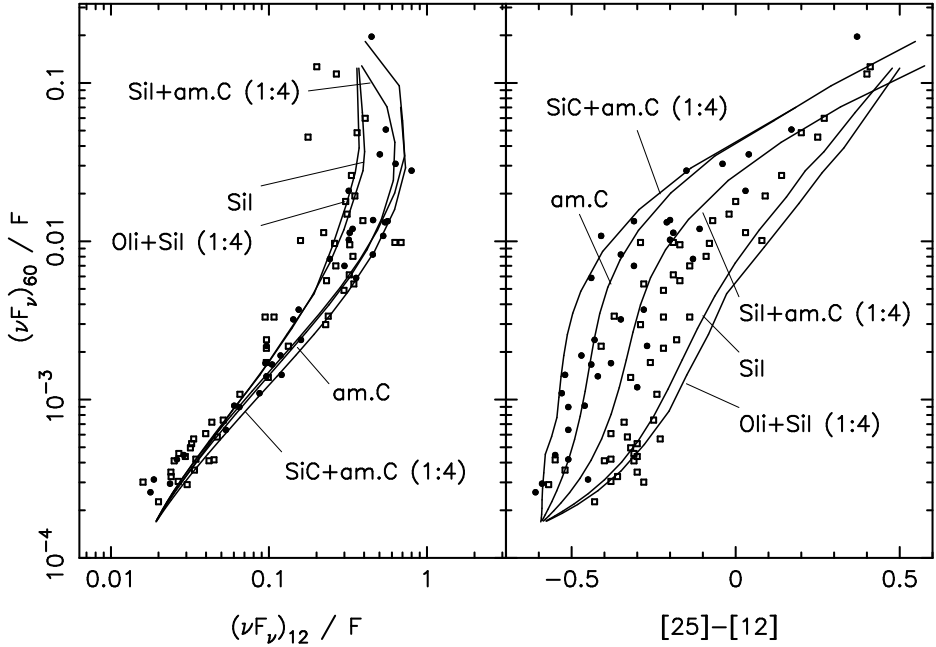}

\caption {Test of correlations predicted by scaling. Open squares mark the
data points for O-stars, solid circles C-stars. Curves describe the model
predictions for various grain compositions, as marked.}

\end{figure}

\section{CONSEQUENCES}

Scaling implies that for a given type of dust, the spectral energy
distributions (SEDs) of late-type stars are controlled almost exclusively by
overall optical depth and thus form a one-parameter family.  Therefore, a
single point on the normalized spectral shape $f_\lambda$ determines the
entire function and any two spectral properties should be correlated with each
other.  Figure 1 displays in its left panel the distribution of normalized
fluxes at 12 and 60 \mic\ for a sample of 89 stars.  These are all the IRAS
objects identified as late-type stars for which we were able to find listings
in the literature for both total fluxes, luminosities, terminal velocities and
mass-loss rates.  As expected from scaling, the normalized fluxes do display a
tight correlation that holds over the entire observed range, covering more
than three orders of magnitude.  The spread can be attributed to the
differences in dust properties between carbon- and oxygen-rich stars.  The
right panel displays the corresponding distribution of $(\nu F_{\nu})_{60}/F$
and [25]--[12] color.  The expected correlation again is evident, the
separation between C and O stars is more pronounced.  The lines are our
computed model spectra, displaying a close agreement with the data for various
relevant chemical compositions (Sil stand for astronomical silicate; am.C for
amorphous carbon; Oli for olivine).  The agreement between model predictions
and observations is better than factor 2 in almost all cases.  In addition,
O-stars and C-stars clearly separate and congregate in accordance with the
trend of the model curves for corresponding chemical compositions.  This
behavior is particularly prominent in the right panel.

\begin{figure}[ht]
\centering \leavevmode
\epsfxsize=0.9\hsize \epsfbox[55 30 560 705]{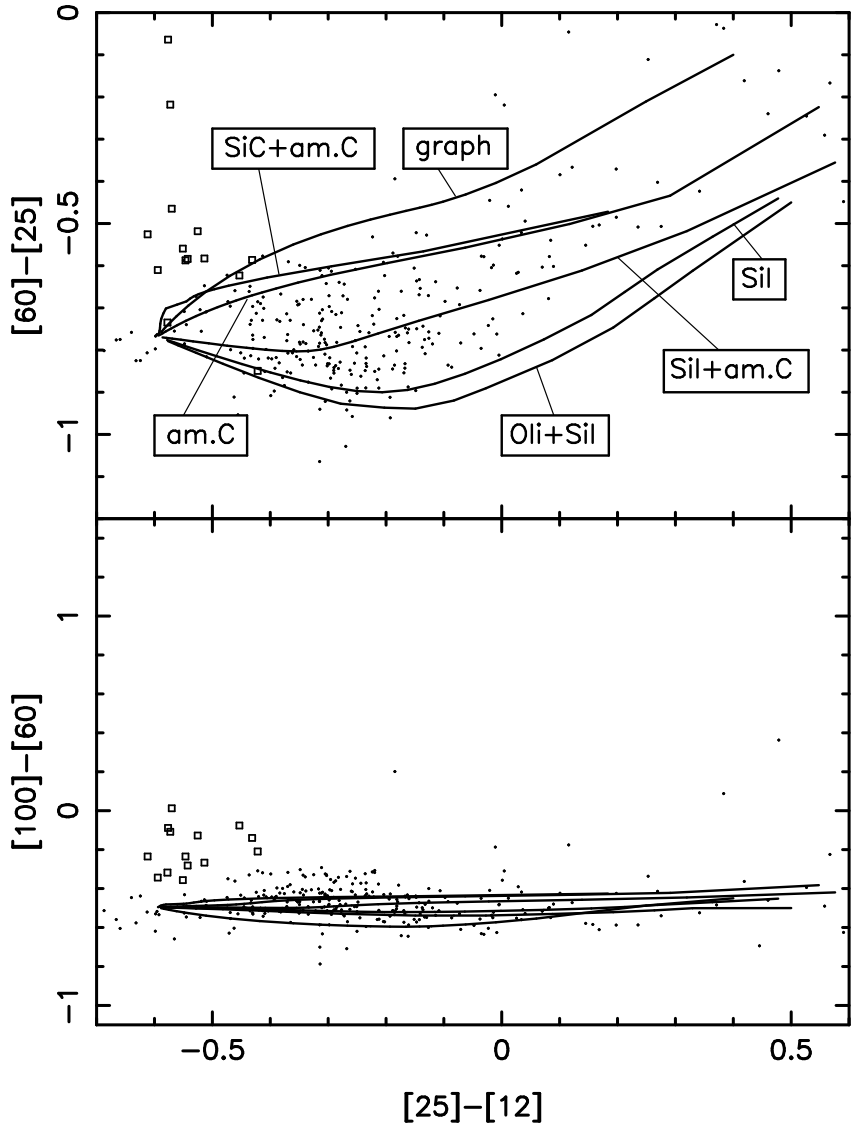}

\caption {IRAS color-color diagrams of all sources in the VH-window with
reliable data.  Solid lines represent theoretical sequences for different
grain compositions, as marked. In all mixtures, the abundances ratio of the
first to second component is 1:4.  In the lower panel, the order of the tracks
from bottom to top around [25]--[12] \about\ 0 is graphite, Oli. + Sil, Sil,
Sil + am.C, am.C and SiC + am.C.}

\end{figure}

Location in IRAS color-color diagrams has become a widely used indicator of
the nature of a source.  Van der Veen \& Habing (1988) identified the
appropriate IRAS region for late-type stars as [60]--[25] $<$ 0 and [25]--[12]
$<$ 0.6, which we will refer to as the {\it VH-window}.  Figure 2 displays the
color-color diagrams for all IRAS sources with reliable data whose colors fall
in this window. Our model predictions for selected grain compositions are
displayed as tracks.  Position along each track is determined by $\tau^T$ and
all tracks originate from the same spot, $\tau^T$ = 0, corresponding to a
black-body spectrum at 2500 K convoluted with the IRAS instrumental profiles.
Distance from this common origin along each track increases with $\tau^T$. As
$\tau^T$ increases, the dust emission becomes more prominent and the tracks of
different grains branch out.  The tracks for purely Si- and C-based grains
outline the distribution boundaries of most IRAS sources in the VH-window,
verifying the VH conjecture that this is the color-color location of late-type
stars.  The data points fill the entire region between these tracks and if
this spread is real, it requires a similar spread in the optical properties of
the grains.  This could reflect chemical mixtures, as we proposed in IE95, or
perhaps grain impurities.

It is important to note that we have removed from our sample sources whose
IRAS fluxes are severely contaminated by cirrus emission.  This contamination
is measured by the cirrus contamination index
\eq{
                  CCI \equiv {\hbox{cirr3} \over F(60)},
}
where cirr3 is the 100 \mic\ cirrus flux at the location of the source and
$F(60)$ is its listed 60 \mic\ flux.  In IE95 we show that the [100]--[60]
color and $CCI$ are perfectly correlated in sources with $CCI \ga 2$,
indicating that the IRAS 60 and 100 \mic\ fluxes of these sources are
unreliable.  Sources in the upper-left corner of the color-color diagrams were
thought to provide evidence for cool shells (Willems \& de Jong 1986), but
almost all of these sources turn out to be cirrus contaminated. Of the total
292 sources displayed in figure 2, only 11 late-type stars (4\%; marked with
open squares) can be identified as a group whose colors are not well explained
by steady-state radiatively-driven winds.  Furthermore, among these 11 sources,
10 have borderline contamination with $1 < CCI < 2$. Based on new simultaneous
observations in a wide spectral range, Miroshnichenko et al. present in a
poster paper model fits for 4 carbon stars. The results are reproduced in
figure 3, showing an excellent agreement with the model for steady-state winds
without any additional shells.  The only parameter that varies from model to
model is the visual optical depth $\tau_v$.  Two of the displayed stars, RY Dra
and VY UMa, belong to the group marked with squares in figure 2. They are
borderline contaminated and show a 100 \mic\ excess proportional to $CCI$, so
clearly they do not give any indication of detached shells.  There is only one
clearly uncontaminated C-star, U Ant, in the ``cool shell" zone.  Indeed, this
star has a thin CO shell (Olofsson et al. 1990).  We conclude that, because of
potential cirrus contamination, the SED is unfortunately not a good indicator
of detached shells.

\begin{figure}[ht]

\centering \leavevmode \epsfxsize=\hsize \epsfbox[15 60 590 770]{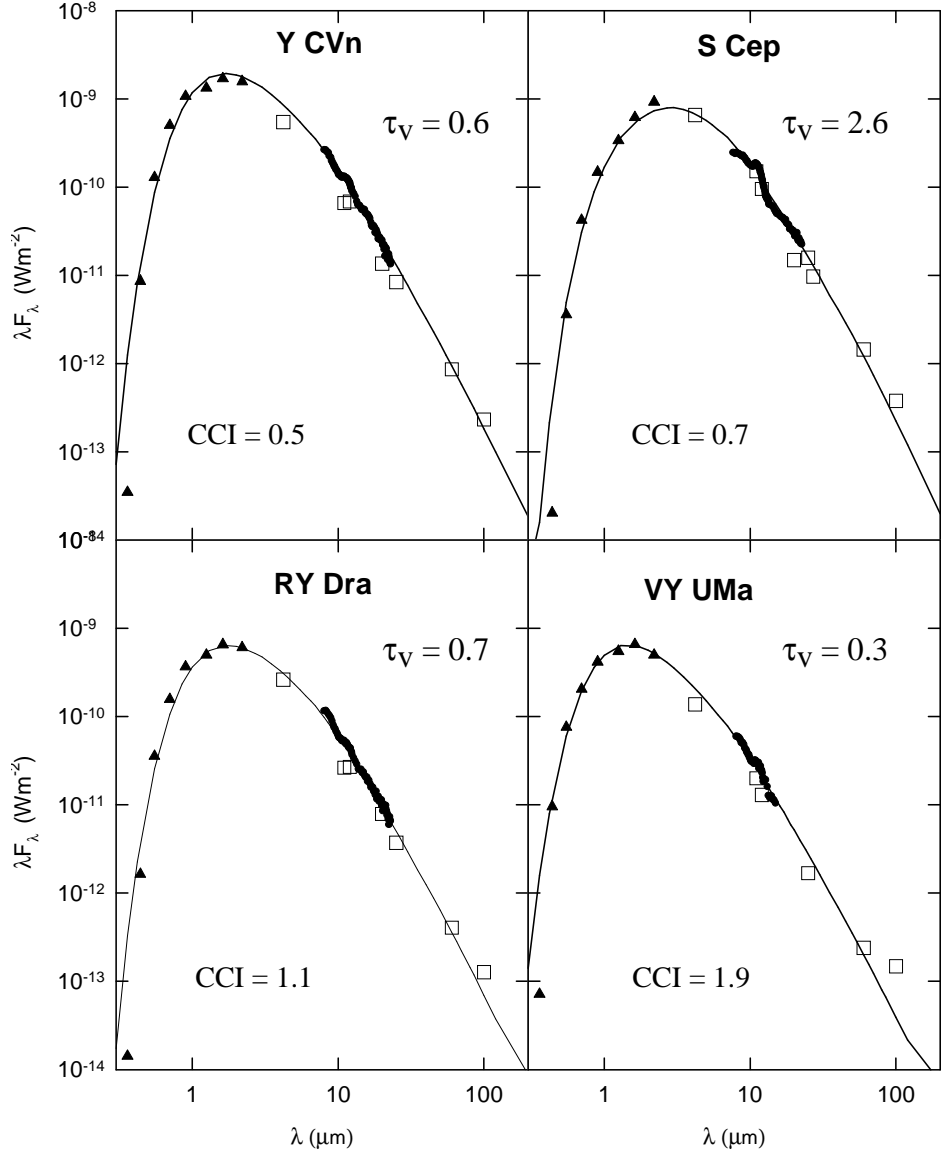}

\caption {Model fits for four carbon stars with new ground based observations
(triangles; Miroshnichenko \& Kuratov 1996).  Circles denote LRS data, squares
are IRAS PSC and RAFGL data. The only adjusted model parameter is the visual
optical depth $\tau_v$.}

\end{figure}

Thanks to scaling, the SEDs of large samples can be modeled systematically. We
have embarked on a large-scale modeling project in which we are fitting all IR
data of all the late-type stars in the IRAS catalogue (\about\ 5,000
sources).  The outcome of this project will be a supplemental catalogue,
listing the optical depths and dust properties of the largest sample yet
modeled with a single, consistent approach.

\section{CONCLUSIONS}

The structure displayed in infrared color-color diagrams is a result of
general scaling properties of radiatively heated dust.  Sources segregate into
families that differ from each other because of dust properties, not because
of the central source.  Within family, location of each source is controlled
by its optical depth.  Concerning late-type stars --- properly selected IRAS
data is almost fully explained by steady-state winds.  The  spectral shapes of
sources free of cirrus contamination do not give evidence for detached shells
in more than a handful of sources.

Support by NASA grant NAG 5-3010, NSF grant AST-9321847 and the Center for
Computational Sciences at the University of Kentucky is gratefully
acknowledged.

\end{document}